\documentclass[%
 reprint,
 twocolumn,
superscriptaddress,
 amsmath,amssymb,
 aps,
prb,
]{revtex4-1}

\usepackage{graphicx}
\usepackage{dcolumn}
\usepackage{bm}
\usepackage{xcolor}

\newcommand{\p}{\partial}
\newcommand{\px}{\partial_x}
\newcommand{\py}{\partial_y}

\begin{document}

\title{Polariton gap and gap-stripe solitons  in Zeeman lattices}

\author{ Dmitry A. Zezyulin}%
\email{dzezyulin@itmo.ru}
\affiliation{%
ITMO University, St.~Petersburg 197101, Russia
}%

\author{Yaroslav V. Kartashov}
\affiliation{%
Institute of Spectroscopy, Russian Academy of Sciences, Troitsk, Moscow, 108840, Russia
}%

\author{Ivan A. Shelykh}
\affiliation{%
ITMO University, St.~Petersburg 197101, Russia
}%
\affiliation{Science Institute, University of Iceland, Dunhagi 3, IS-107, Reykjavik, Iceland}

\date{\today}

\begin{abstract}
We predict that spatially modulated Zeeman splitting resulting in the formation of Zeeman lattice can be used for creation of localized self-sustained excitations in spinor polariton condensates with dominant repulsive interactions. In such lattices, the phenomenon of TE-TM splitting, playing the role of effective spin-orbit interaction, leads to the emergence of the stripe phase and formation of stable gap-stripe solitons with complex intrinsic structure resulting from the presence of two characteristic spatial scales, one of which is set by the period of Zeeman lattice, while other is set by the momentum in the depth of the Brillouin zone, at which such solitons bifurcate from the linear spectrum. Gap-stripe polariton solitons can be excited by suitable resonant pump.
\end{abstract}


\maketitle

\section{Introduction}
Magnetic fields can qualitatively change the properties of physical systems, and in some cases even lead to the appearance of the new states of matter. {Examples include incompressible Fermi liquid in the regime of the fractional quantum Hall effect \cite{QHE}, destruction of superconductivity by magnetic fields exceeding critical value \cite{Superconductivity}, magnetic field induced crossover between regimes of weak localization and anti-localization \cite{Altshuler}, and others}. In the systems consisting of neutral particles, such as cold atoms, excitons or cavity polaritons, magnetic field can not influence the orbital motion directly. However, it can still strongly affect characteristics of the system by acting on the spin of the particles. In this context, polaritonic systems reveal particularly rich phenomenology.

Cavity polaritons (also known as exciton-polaritons) \cite{Hui2010} are hybrid light-matter quasiparticles appearing in the regime of strong coupling between confined mode of a planar microcavity and an excitonic transition brought in resonance with it. Polaritons possess a set of peculiar properties which make them ideal candidates for observation of quantum collective phenomena at surprisingly high temperatures \cite{Carusotto2013}. Many of them are related with spin structure of polaritons, inherited from spin structure of individual excitons and photons \cite{ShelykhReview}. 

Polariton spin can be affected by application of an external magnetic field, which results in the Zeeman splitting in circular polarized components of a polariton doublet and by effective magnetic fields of various origin. Among these latter, one should mention TE-TM splitting of the photonic modes of a planar cavity, which results in the appearance of k-dependent in-plane effective magnetic field. Its role is similar to that played by spin-orbit interaction in electronic systems. Moreover, spin anisotropy of polariton-polariton interactions \cite{Glazov2009,Sich2014} gives rise to the onset of an additional magnetic field directed along the structure growth axis and dependent on the polariton concentration and polarization. The interplay between the effects caused by real and effective magnetic fields results in a plethora of intriguing phenomena, such as spin Meissner effect \cite{Rubo2006,Larionov2010,Gulevich2016}, generation of synthetic gauge fields \cite{Tercas2014,Nalitov2015,Zezyulin2018,Shelykh2018} and formation of nontrivial topological phases in polariton lattices \cite{Nalitov2015a,Solnyshkov2016,SkrKar2016,Klembt2018,SkrKar2019}. 

In this paper we predict a different phenomenon, which appears due to the interplay between periodically modulated Zeeman splitting, TE-TM splitting, and spin-anisotropic polariton-polariton interactions. We consider a system, where the value of the Zeeman splitting oscillates periodically along one axis. Experimentally, this can be realized by application of the inhomogeneous magnetic field,   by patterning of a semimagnetic cavity for which polariton g-factor is dramatically enhanced \cite{Rousset2017,Mirek2017,Krol2019}, {by incorporating magnetic ions \cite{Mietki2018} and by microcavity etching \cite{Sun2019}. In atomic condensates, periodic  Zeeman lattice (ZL)  has been experimentally synthesized   using a combination of applied radio-frequency magnetic   and Raman fields that simultaneously couple the atomic spin states \cite{Jimenez}}. We demonstrate that 
the periodic  ZL
enables the formation of a rich variety of stable gap polariton solitons. Moreover, under appropriate conditions  TE-TM splitting induces stripe phase, leading to the appearance of gap-stripe solitons with complex internal structure bifurcating from the internal points of the reduced Brillouin zone (BZ) and having no counterparts in conventional polaritonic lattices \cite{lattpolariton1,lattpolariton2,lattpolariton3,lattpolariton4,lattpolariton5,lattpolariton6}. Our results thus open the way for experimental realization of such previously elusive topological defects in a unique setting {which exposes the spinor components to opposite effective potentials and hence results in physics distinctively different from that in better studied conventional optical lattices \cite{OL}, where both components are subjected to equal potentials.}
%
%

The rest of the paper is organized as follows. In Sec.~\ref{sec:linear} we introduce the physical model and study its band-gap structure in the linear limit. In Sec.~\ref{sec:stripe} we present gap-stripe solitons that form in Zeeman lattices with weak modulation. In Sec.~\ref{sec:deep} and Sec.~\ref{sec:nonzero} the consideration is extended on the case of deep Zeeman lattices and lattices with nonzero mean, respectively. In Sec.~\ref{sec:2D} we address dynamics of two-dimensional solitons. Section~\ref{sec:concl} concludes the paper.

\begin{figure*}[t]
\begin{center}
\includegraphics[width=0.999\textwidth]{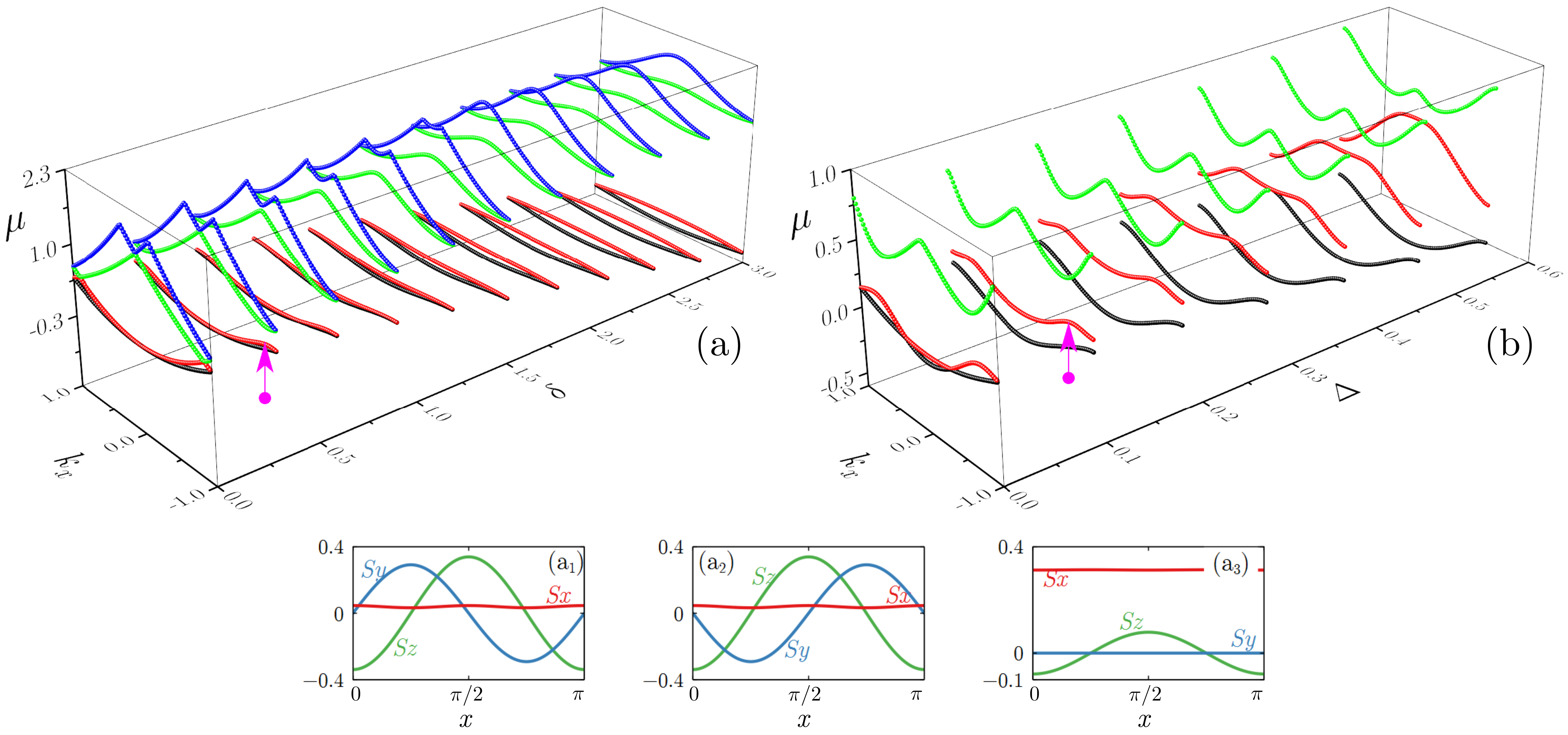}
\caption{(a) Transformation of the band‐gap structure of the ZL with zero mean   $\Delta= 0$ for $\beta=\pm  0.1$ and gradually increasing lattice depth $\delta$. {Stokes parameters for $\delta=0.3$ for the stripe phase with $k_x\approx 0.92$ (corresponding to the magenta arrow) (a$_1$), $k_x \approx -0.92$ (a$_2$), and in the center of the BZ $k_x=0$ (a$_3$). } (b) Transformation of the band‐gap structure  for $\beta=\pm 0.3$, lattice depth $\delta  = 0.35$, with increase of the mean level $\Delta$. Magenta arrows indicate some of $k_x$ values at which stripe solitons emerge. }
\label{fig:bands}
\end{center}
\end{figure*}

\section{Model and linear dispersion relations}
\label{sec:linear}
The evolution of spinor polariton wavefunction ${\bf \Psi} = (\psi_+, \psi_-)^\textrm{T}$ in circular polarization basis is governed by the dimensionless Gross-Pitaevskii equations (see e.g.   \cite{Flayac2010,Gulevich2016} and Appendix~\ref{app:1}):
 \begin{eqnarray}
   i\p_t \psi_\pm = \left[-\frac{1}{2}(\p_x^2+\p_y^2) + |\psi_\pm|^2+\sigma|\psi_{\mp}|^2 \pm \Omega(x)\right]\psi_\pm\nonumber\\[2mm]
    +\beta(\p_x \mp i\p_y)^2\psi_\mp -i\gamma\psi_\pm + H_{\pm}(x,y)e^{-i\varepsilon t}.
    \label{eq:Omega(x)}\hspace{2cm}
    \label{eq:full}
\end{eqnarray}
Here $x,y$ are the spatial coordinates; $t$ is time; $\beta$ is the spin-orbit coupling (SOC) coefficient  resulting from different effective masses of TE and TM polaritons; the condensate is dominated by strong repulsion between polaritons with the same spin, while  $\sigma=-0.05$ accounts for weak attraction between spin-positive and spin-negative polaritons; $\gamma$ is the polariton loss coefficient; the terms $\sim H_\pm$ describe resonant pumping with frequency detuning $\varepsilon$. Assuming the effective polariton mass $m^*\approx 10^{-34}$\,kg and the unit length $\ell \approx 1\,\mu$m, time unit in Eq.~(\ref{eq:full})  is  $\tau=m^*\ell^2/\hbar\approx 1$\,ps, while characteristic energy is ${\cal E}  = \hbar^2/(m^*\ell^2) \approx 0.7$~meV. Zeeman lattice  $\Omega(x)=(2{\cal E})^{-1}g\mu_\textrm{B} B(x)$ results from spatially modulated applied magnetic field $B(x)$, where $g$ is the effective exciton-polariton $g$-factor and $\mu_\textrm{B}$ is Bohr magneton. We model ZL using $\pi$-periodic function $\Omega(x) = \Delta + \delta  \cos(2x)$, where $\Delta$ and $\delta$ describe constant and spatially modulated constituents of the Zeeman splitting, respectively.

Since we are   interested in the possibility to use ZL for localization of nonlinear polariton states in the $x$ direction, we assume that the polariton condensate is uniform in the $y$ direction, i.e., $\partial_y=0$. To identify possible types of   solitons in the microcavity, we start from the conservative limit by setting $\gamma=0$ and $H_\pm\equiv 0$ in (\ref{eq:full}). The domain of soliton formation is determined by the band-gap structure of the underlying linear ZL which is obtained when nonlinear terms $\sim|\psi_\pm|^2, |\psi_\mp|^2$ are neglected. Linear Bloch waves in the resulting periodic system have the form $\psi_\pm = e^{-i\mu(k_x)t + ik_x x} U_\pm(x)$, where $U_\pm(x)$ are $\pi$-periodic functions, $k_x\in(-1, 1]$ is the quasimomentum in the reduced BZ. Lowest allowed bands $\mu(k_x)$ computed numerically \cite{ffhm} for the ZL with zero and nonzero mean   $\Delta$ are shown in Fig.~\ref{fig:bands}(a) and (b), respectively. Starting from the $\Delta=0$ case, we observe that at zero  lattice depth $\delta$ [see Fig.~\ref{fig:bands}(a)], the spectrum consists of two folded (and hence intersecting at nonzero $k_x$) parabolas $\mu=k_x^2(1/2\pm \beta)$ and is gapless, except for the semi-infinite gap $\mu \in (-\infty, 0]$, where bright solitons cannot exist due to polariton-polariton repulsion. When the depth $\delta$  is nonzero, a finite gap opens in the vicinity of the intersections between two parabolas. As a result, maxima (minima) of the lower (upper) dispersion curves are achieved in the internal $k_x$ points of the Brillouin zone [magenta arrow in Fig.~\ref{fig:bands}(a)]. {Thus, the combination of SOC and spatially modulated Zeeman splitting leads to stripe phase, akin to that in atomic systems with pseudo-SOC \cite{stripe1,stripe2}, that can potentially result in gap-stripe solitons \cite{free1dim1,perlattice} when condensation in the Zeeman lattice is achieved in the nonlinear regime, e.g. in the presence of the external pump \cite{dark13}.} 
Increasing $\delta$ broadens the  gap, but  inhibits stripe phase by shifting extrema of dispersion curves towards $k_x=0$ or $1$.    {Peculiarity of the stripe phase becomes evident from the  pseudo-spin texture described by the Stokes parameters $S_{\{x,y,z\}} = (\mathbf{\Psi}^\dag \sigma_{\{x,y,z\}}\mathbf{\Psi})/(\mathbf{\Psi}^\dag\mathbf{\Psi})$, where $\sigma_{x,y,z}$ are   Pauli matrices, and $^\dag$ is Hermitian conjugation.
Spin texture for the stripe phase, shown in Fig.~\ref{fig:bands}(a$_{1,2}$) for positive and negative $k_x$-values, is more complex than that in the center of BZ at $k_x=0$ shown in Fig.~\ref{fig:bands}(a$_{3}$). In the latter case the diagonal Stokes parameter is identically zero: $S_y=0$, whereas spin texture of the stripe phase is characterized by nontrivial distributions of all Stokes components.}

\section{Gap-stripe solitons in weak Zeeman lattices with zero mean}
\label{sec:stripe}
\begin{figure}
\begin{center}		 
\includegraphics[width=0.99\columnwidth]{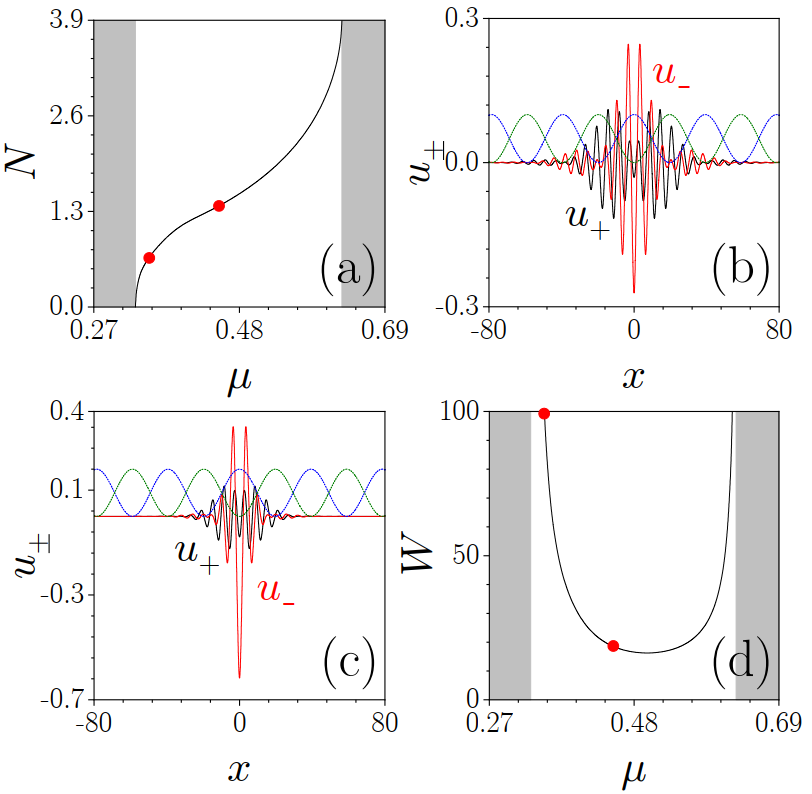}
\caption{Norm $N$ (a) and width $W$ (d) of the gap‐stripe solitons versus chemical potential $\mu$ in the ZL with zero mean $\Delta=0$ at $\delta=0.3$ and $\beta =0.1$. Gray shaded domains indicate allowed spectral bands. Examples of gap-stripe solitons at $\mu=0.35$ (b) and $\mu=0.45$ (c) corresponding to the red dots in (a), (d), that are weakly and strongly localized, respectively. Dashed curves in (b),(c) are proportional to $\sin^2[(1-\tilde{k}_x)x]$ and $\cos^2[(1-\tilde{k}_x)x]$, where $\tilde{k}_x\approx 0.92$, and highlight new spatial scale emerging from the gap‐stripe phase.}
\label{fig:shallow}
\end{center}
\end{figure}

Now we turn to the nonlinear case and study polariton solitons in the finite first gap of the ZL that opens at $\delta>0$. Considering high-finesse microcavity we first address such states in conservative system with $\gamma,H_{\pm}=0$. Stationary states have the form $\psi_\pm = e^{-i\mu t} u_\pm(x)$, where real $\mu$ is the chemical potential, and $u_\pm$ are localized: $|u_\pm|\to 0$ as $|x|\to\infty$. The central result of this paper is the existence of localized stripe-gap solitons bifurcating from lower gap edge at quasimomentum values $\pm \tilde{k}_x$ in the internal points of the reduced Brillouin zone, whose properties are summarized in Fig.~\ref{fig:shallow}. In comparison with conventional gap solitons, gap-stripe solitons feature more complex internal structure, because in the vicinity of the bifurcation their form is determined by a superposition $U_\pm(x)e^{i\tilde{k}_x x} + U_\pm^*(x)e^{-i\tilde{k}_x x}$ of two Bloch states with slowly decaying envelope. In the particular case of weak ZL with $\delta=0.3$ and zero mean $\Delta$ the bifurcation of gap-stripe soliton occurs at $\tilde{k}_x \approx \pm 0.92$ for $\beta=0$. Therefore   in the vicinity of the bifurcation, under broad decaying envelope of the gap-stripe soliton one observes not only fast oscillations with a period equal to that of the lattice, but also slow modulation with characteristic spatial scale $\pi/(1-\tilde{k}_x)\approx 40$, clearly visible in Fig.~\ref{fig:shallow}(b). To characterize the family of the gap-stripe solitons, we introduce total norm of the spinor wavefunction $N = \int_{-\infty}^\infty(|u_+|^2 + |u_-|^2)dx$, which  increases with increase of the peak condensate density/amplitude, and soliton integral width $W = N^{-1}\int_{-\infty}^\infty (x-X)^2 (|u_+|^2 + |u_-|^2)dx$, where $X = N^{-1} \int_{-\infty}^\infty x (|u_+|^2 + |u_-|^2)dx$ is the center of mass. The gap-stripe soliton family emerging at the lower gap edge at small amplitudes, vanishes at its upper edge [Fig.~\ref{fig:shallow}(a)], where soliton's amplitude remains finite. The width of soliton plotted in Fig.~\ref{fig:shallow}(d), being nonmonotonic function of $\mu$, diverges at both edges of the gap. The example of strongly localized soliton with nearly minimal width is shown in Fig.~\ref{fig:shallow}(c).

\begin{figure*}
\begin{center}
\includegraphics[width=0.99\textwidth]{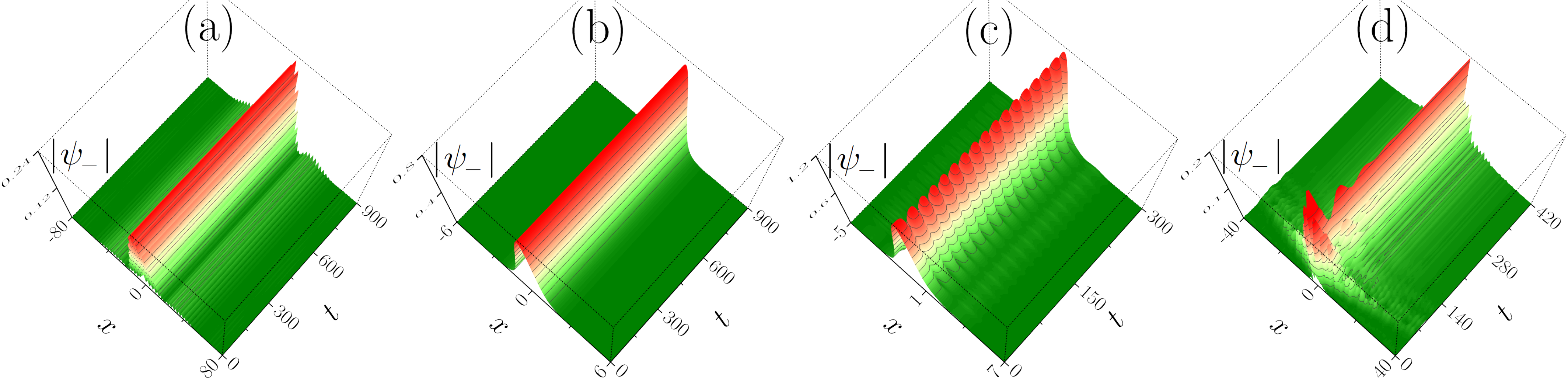}
\caption{Stable evolution of (a) stripe‐gap broad soliton with $\mu= 0.35$ in shallow ZL with $\delta=0.3$ and of (b) strongly localized
symmetric soliton with $\mu=-1$ in deep ZL with $\delta=3$ . (c) Formation of persistent breather from unstable soliton with shifted in-phase components  at $\mu=-0.34$, $\delta=3$.  In (a,b,c)  $\beta=0.1$, $\Delta=0$. (d)  Excitation of a stripe soliton in the second gap in ZL with nonzero mean using resonant pump. Here $\Delta=0.45$, $\delta=0.3$, $\beta=0.3$, $\varepsilon=0.41$. Initial condition is small-amplitude random noise. In all cases  only $\psi_-$ component is shown. }
\label{fig:dynamics}
\end{center}
\end{figure*}

We examine soliton stability using linear stability analysis \cite{JY}, i.e.,  introduce a perturbed solution   $\psi_\pm = e^{-i\mu t}[u_\pm(x)+\xi_\pm(x)e^{i\lambda t} + \chi_\pm(x)e^{-i\lambda^* t}]$, linearize governing equations with respect to small perturbations $\xi_\pm$,   $\chi_\pm$, and evaluate the spectrum of exponents $\lambda$. For stable solitons all  $\lambda$ should be real. Stability analysis demonstrates that gap-stripe soliton family from Fig.~\ref{fig:shallow} is stable for small and moderate values of the norm $N$. Stability has also been validated using direct simulations of the soliton evolution up to $t\sim 10^4$ with small random complex initial perturbations. Representative example of stable evolution is shown in Fig.~\ref{fig:dynamics}(a). For sufficiently large norms $N$ gap-stripe solitons become unstable with  weak oscillatory instabilities ubiquitous for repulsive condensates \cite{instability}. However, the increments of such instabilities (i.e., imaginary parts of eigenvalues $\lambda$) may be very small, so that instability development sometimes requires huge times in comparison with polariton lifetime. On this reason, we do not mark exact stability border in Fig.~\ref{fig:shallow}, but mention that appreciable instabilities are detected for solitons with $\mu>0.5$.

\section{Solitons in deep Zeeman lattices}
\label{sec:deep}
\begin{figure}
\begin{center}	
\includegraphics[width=0.99\columnwidth]{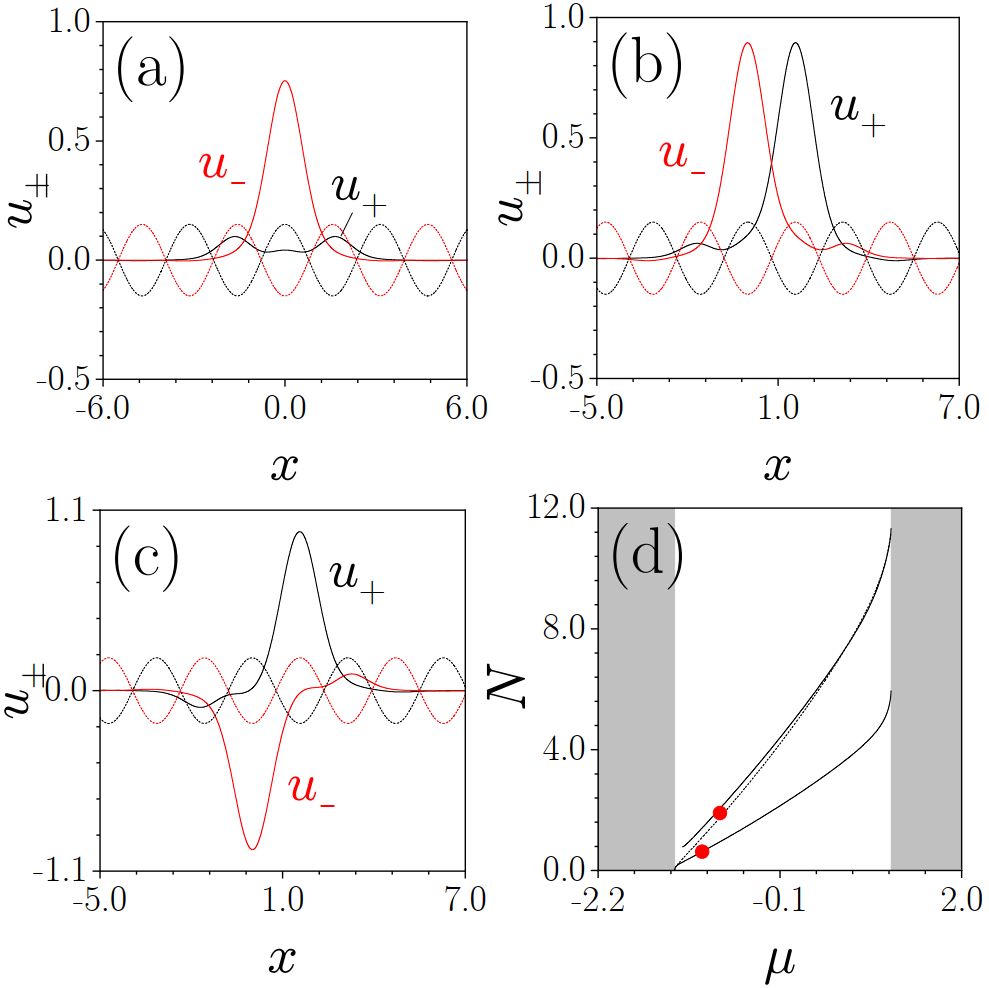}
\caption{(a) Profile of the simplest symmetric soliton in deep ZL with $\delta=3$ at $\mu=-1$, and profiles of dipole-like solitons with in‐phase (b) and out‐of‐phase spatially shifted components at $\mu=-0.79$. Dashed lines schematically indicate ZL profile for $u_+$ and $u_-$ components. (d) Norm $N$ versus chemical potential  $\mu$ for above soliton families (lower solid line corresponds to symmetric soliton, while upper solid and dashed lines correspond to out-of-phase and in-phase dipoles, respectively). Red dots correspond to profiles  in (a)‐(c).}
\label{fig:deep}
\end{center}
\end{figure}

We proceed to solitons in ZL with larger depths $\delta$. According to  Fig.~\ref{fig:bands}(a), increasing $\delta$ inhibits stripe phase in the lowest finite gap, but 
increases the width of this gap, thereby substantially enriching the variety of the coexisting stable solitons. The simplest gap soliton in deep ZL has symmetric profile in both  components, $u_\pm(x)=u_\pm(-x)$, see Fig.~\ref{fig:deep}(a). This soliton coexists with the unstable in-phase [Fig.~\ref{fig:deep}(b)] and stable out-of-phase [Fig.~\ref{fig:deep}(c)] dipole-like solitons with shifted peak locations in spin-positive and spin-negative components. Notice that such solitons have never been considered in polariton condensates and they exist only in ZL, where two components feel opposite potentials and, hence, tend to populate spatially shifted minima of respective potentials. Soliton families $N(\mu)$ presented in Fig.~\ref{fig:deep}(d)  reveal that symmetric solitons and in-phase dipoles emerge at the left gap edge (from the small norm limit $N\to 0$), whereas out-phase-dipoles exist only if norm $N$ exceeds a nonzero threshold.
In-phase (out-of-phase) dipole solitons are typically unstable (stable) for positive (negative) values of the SOC coefficient $\beta$, which is can be understood from  the contribution of SOC into the total  energy functional (see Appendix~\ref{app:2}). Stability of symmetric and out-of-phase solitons 
has also been confirmed in dynamical simulations. Figure~\ref{fig:dynamics}(b) showcases stable evolution of a representative symmetric soliton. Interestingly, dynamical instability of the in-phase dipoles triggers formation of persistent breathers, whose shape changes periodically upon evolution, see Fig.~\ref{fig:dynamics}(c).

\begin{figure}
\begin{center}	
\includegraphics[width=0.99\columnwidth]{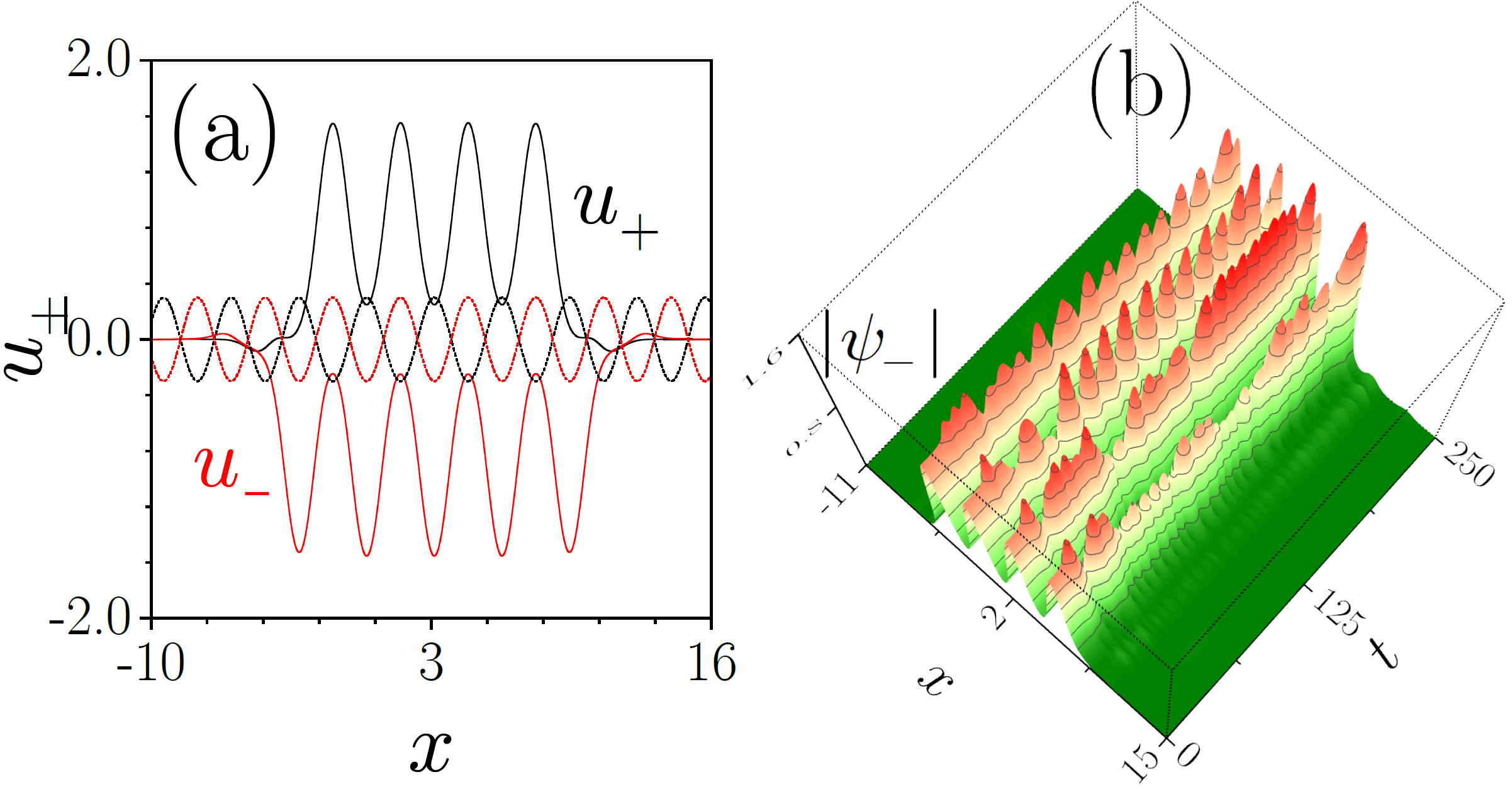}
\caption{(a) Example of a stable out-of-phase gap soliton train in deep ZL at $\mu=0.25$. (b) Decay of an unstable in-phase soliton train at $\mu=-0.34$. Only $\psi_-$ component is shown. In both panels  $\delta = 3$, $\beta=0.1$, $\Delta=0$.}
\label{fig:trains}
\end{center}
\end{figure}

Another distinctive class of nonlinear localized states in ZL corresponds to soliton trains, i.e., complex multipole structures with several density peaks. These solitons can be interpreted as the truncated nonlinear Bloch waves \cite{trains1,trains2}. There exists a rich variety of  such states, with equal or different number of peaks in each component. Similar to their dipole counterparts, in-phase (out-of-phase)   soliton trains are typically unstable (stable) for positive (negative) values of the SOC coefficient $\beta$.  Representative stable out-of-phase soliton train with different numbers of peaks in $\psi_{\pm}$ components is shown in Fig.~\ref{fig:trains}(a), while Fig.~\ref{fig:trains}(b) presents typical evolution dynamics of the unstable soliton train. All of such states typically exist above minimal norm $N$.

\section{Solitons in Zeeman lattices  with nonzero mean} 
\label{sec:nonzero}
\begin{figure}
\begin{center}		
\includegraphics[width=0.99\columnwidth]{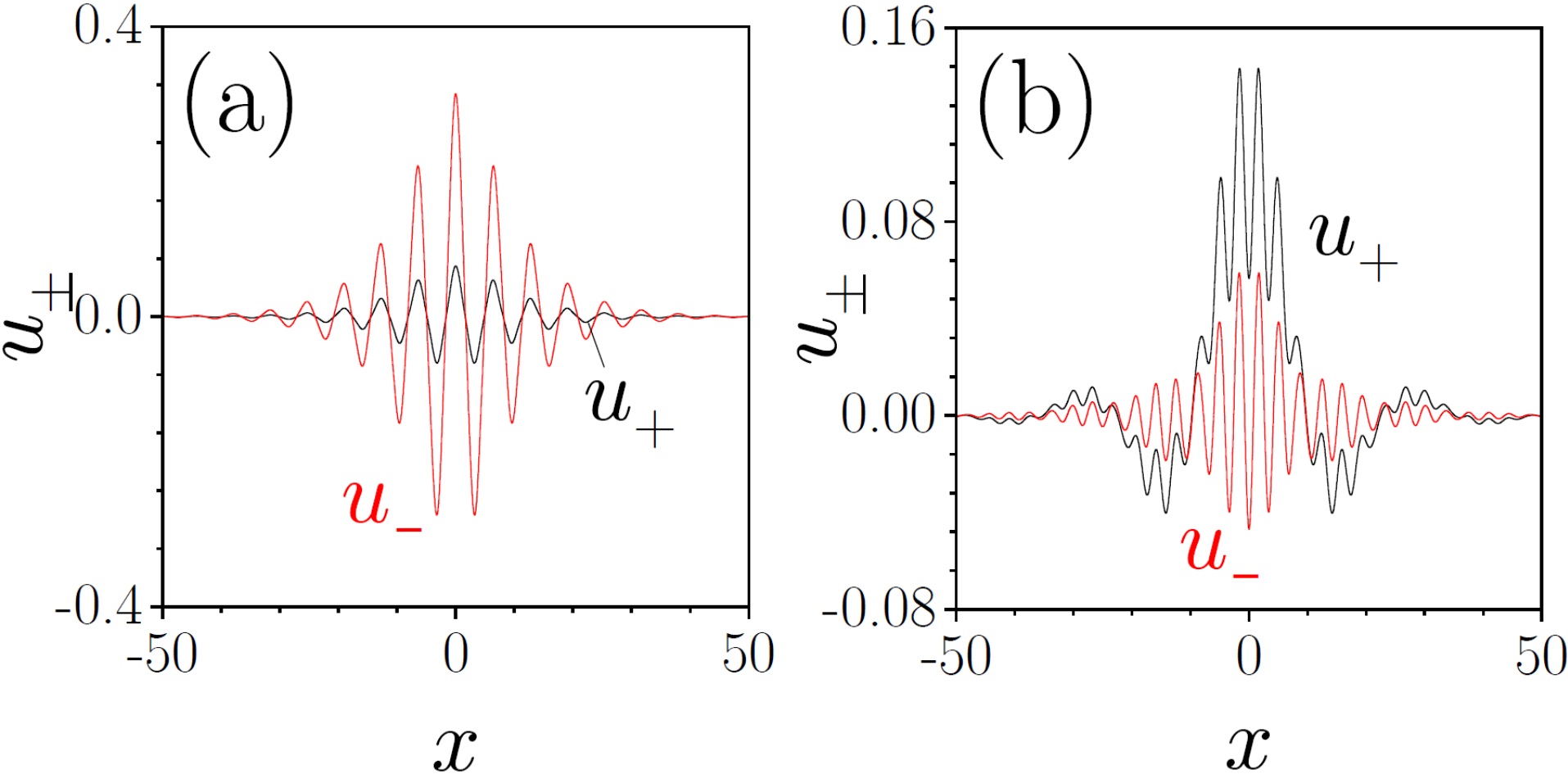}
\caption{Solitons in ZL with nonzero mean. (a) Gap  soliton from the first finite gap with $\mu=-0.145$.   (b) Gap-stripe soliton from the second finite gap at $\mu=0.41$. In both cases $\Delta=0.45$,   $\delta=0.3$, $\beta=0.3$.}
\label{fig:Delta}
\end{center}
\end{figure}

Remarkably, SOC acting in polariton system can induce stripe phases not only in lower, but also in higher gaps of the ZL with nonzero mean $\Delta\ne 0$. This is illustrated in Fig.~\ref{fig:bands}(b), where we show transformation of the bang-gap spectrum with the increase of the mean value $\Delta$. For  
sufficiently strong SOC, gap-stripe solitons 
emerge in the second finite gap [see magenta arrow in Fig.~\ref{fig:bands}(b) indicating $k_x$ at which such solitons bifurcate from the linear spectrum]. Such a lattice then simultaneously supports conventional stable gap solitons [Fig.~\ref{fig:Delta}(a)] in the first gap and gap-stripe solitons in the second gap [Fig.~\ref{fig:Delta}(b)]. 

Polariton condensates are essentially dissipative  and require external pump 
\cite{dark13}. The advantage of the ZL system is that all states reported above can be excited with suitable resonant pump. To illustrate this, we consider the complete nonequilibrium system (\ref{eq:full}) with  typical polariton loss coefficient $\gamma=0.02$ and localized resonant pump $H_\pm=H_0 \exp\{-x^2/w_0^2\}$ with $H_0=0.05$ and $w_0^2=10$. The solitons are efficiently excited when pump frequency detuning $\varepsilon$ is chosen in the forbidden   gap. Gap-stripe solitons can be excited even from noisy random inputs. Example of the resonant excitation of the gap-stripe soliton in ZL with nonzero mean is shown in  Fig.~\ref{fig:dynamics}(d).

\section{Two-dimensional soliton dynamics} 
\label{sec:2D}
\begin{figure}
\begin{center}	
\includegraphics[width=0.99\columnwidth]{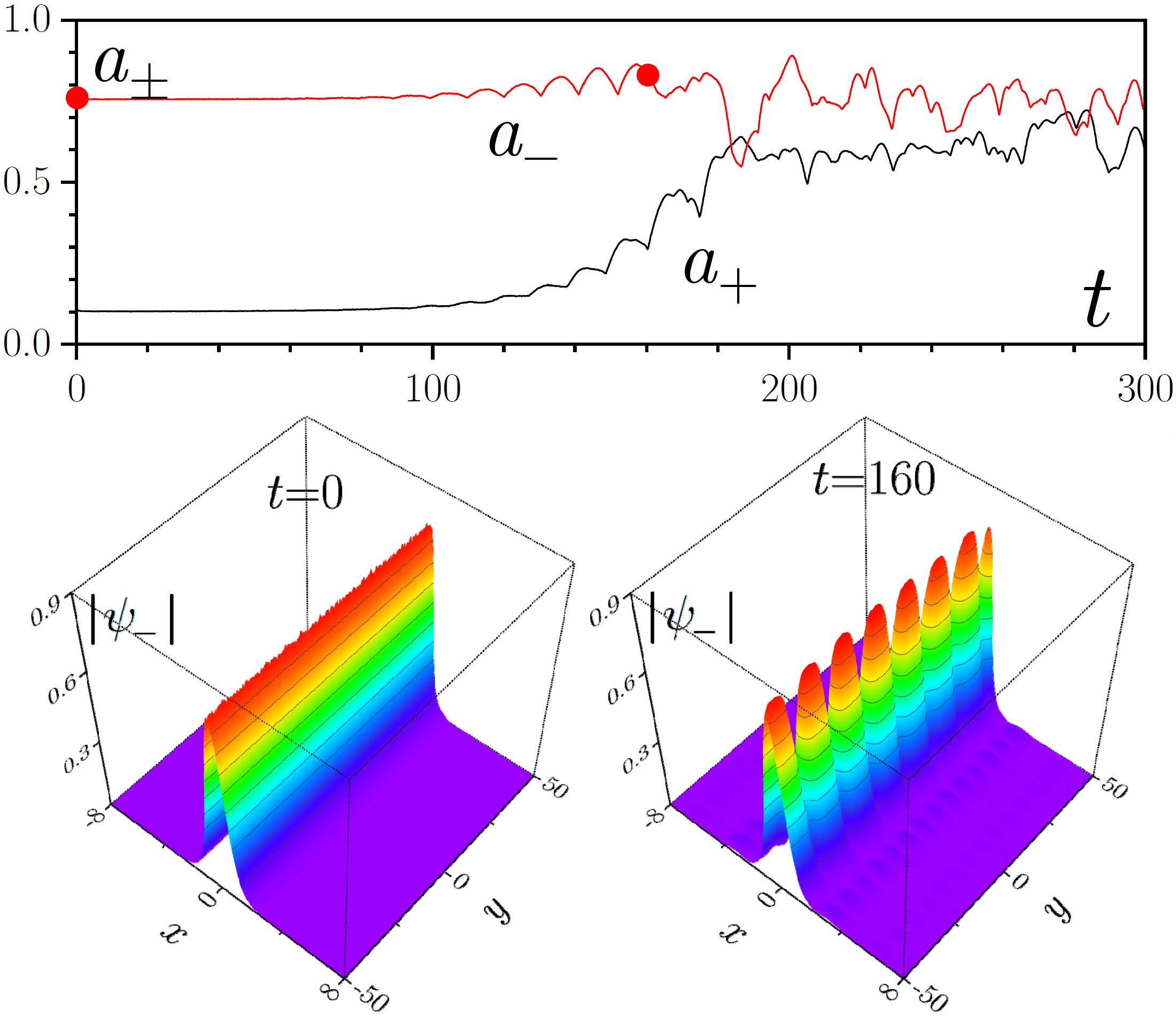}
\caption{Dynamics of development of transverse instability of perturbed symmetric soliton in deep ZL obtained at $\mu=-1$, $\Delta=0$, $\delta =3$ , $\beta =0.1$. Top row
shows peak amplitudes of two soliton components $a_\pm=\max_{x,y}|\psi_\pm|$ \emph{vs} time, while bottom
row show $\psi_-$-distributions in selected moments of time corresponding to the red dots in the top panel.}
\label{fig:S1}
\end{center}
\end{figure}

While the above results demonstrate the formation of stable polariton solitons in the one-dimensional (1D) ZLs, they disregard dynamics in the second spatial dimension $y$, since wavefunctions $\psi_\pm$ are considered $y$-independent. When this dimension is taken into account, full two-dimensional (2D) model in Eq.~(\ref{eq:full}) implies the possibility of the transverse instabilities \cite{TI} of quasi-1D soliton stripes, excluded in the 1D geometry. Such instabilities can also develop in our system despite the fact that polariton-polariton interactions are predominantly repulsive. In Fig.~\ref{fig:S1} we present an example of the transverse instability development for simple quasi-1D (i.e. polariton wavefunctions $\psi_\pm$ in such soliton are initially uniform in the $y$-direction) symmetric soliton from Fig.~\ref{fig:deep}(a). Its evolution dynamics has been modeled using full 2D model in Eq.~(\ref{eq:full}). The instability triggered by small-amplitude input noise results in the development of deep $y$-modulation of the initially uniform soliton stripe.

\begin{figure}
\begin{center}	
\includegraphics[width=0.99\columnwidth]{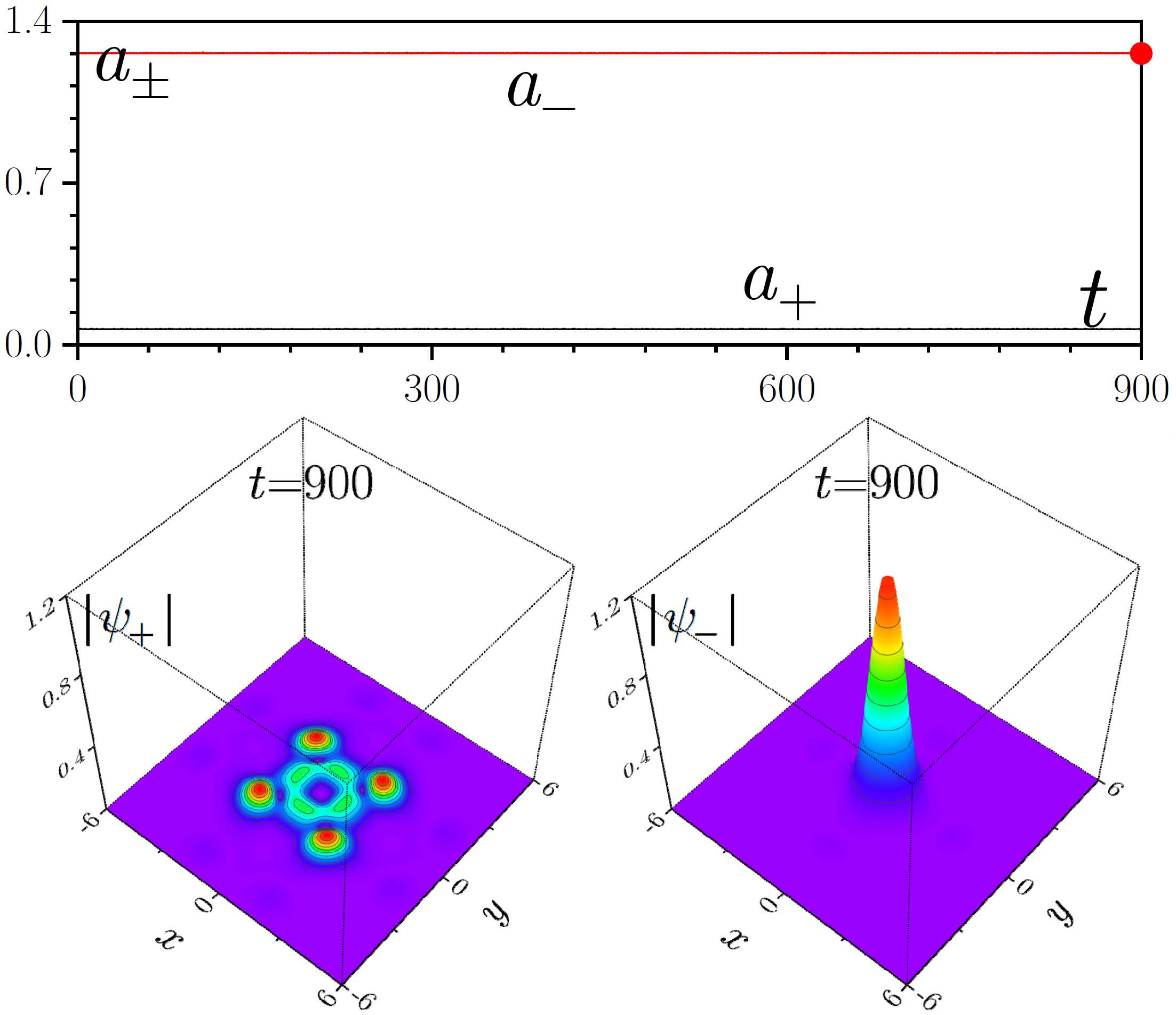}
\caption{Stable evolution of perturbed symmetric soliton obtained at $\mu=-2$, $\Delta=0$, $\delta =3$, $\beta =0.1$ in two‐dimensional ZL. Top row shows peak amplitudes of two
soliton components $a_\pm=\max_{x,y}|\psi_\pm|$  \emph{vs.} time, while bottom row show $\psi_\pm$ distributions at $t =900$.}
\label{fig:S2}
\end{center}
\end{figure}

Such transverse instabilities can be arrested in polariton microcavity wires providing strong confinement in the $y$-direction (such structures were realized experimentally  \cite{2D}). Another promising route to realization of fully 2D stable solitons may be based on 2D Zeeman lattices. In Fig.~\ref{fig:S2} we present an example of stable 2D (i.e., localized in both $x$ and $y$ directions) symmetric gap soliton obtained in the 2D Zeeman lattice $\Omega(x,y) = \Delta  +  \delta[\cos(2x) + \cos(2y)]$ with $\Delta=0$ and $\delta=3$. Stability of such 2D solitons was confirmed by simulation of their evolution dynamics in the presence of small initial noise: corresponding dependencies of peak amplitudes of two soliton components on time are presented in the top row of Fig.~\ref{fig:S2}.   

\section{Conclusion}
\label{sec:concl}
To conclude, we have demonstrated that periodically modulated magnetic field (i.e., Zeeman lattice) enables localization of interacting polaritons with the formation of a variety of gap and gap-stripe solitons. The latter are characterized by complex internal structure, but can be stable. Deep Zeeman lattices support in-phase and out-of-phase dipole solitons, as well as more complex soliton trains with multiple peaks. In Zeeman lattices with nonzero mean, gap-stripe solitons exist in higher spectral gaps and can be excited with resonant pump. When the second spatial dimension is taken into account, quasi-one-dimensional gap solitons can be prone to transverse instabilities. However the latter can be suppressed in two-dimensional Zeeman lattices periodically modulated in both spatial dimensions.

\begin{acknowledgments}
The work of D.A.Z. and I.A.S. is supported by the Government of the Russian Federation through the Megagrant 14.Y26.31.0015, and ITMO 5-100 Program. I.A.S. acknowledges support from Icelandic Science Foundation project ``Hybrid Polaritonics''.  The work of Y.V.K.   was partially supported by the program 1.4 of Presidium of RAS ``Topical problems of low temperature physics''.
\end{acknowledgments}

\appendix 
\section{Normalization of the Gross-Pitaevskii equation}
\label{app:1}
In order to describe the dynamics of a spatially localized polariton condensate, we use a system of Gross-Pitaevskii equations  (for the sake of simplicity here we consider  conservative limit) 
\begin{eqnarray*}
   i\hbar \frac{\p \Psi_\pm}{\partial T}  = -\frac{\hbar^2}{2m^*}\left(\frac{\p^2 \ }{\p X^2} + \frac{\p^2\ }{\p Y^2}\right)\Psi_\pm \pm \frac{g\mu_B}{2}B(X) \Psi_\pm \nonumber\\[3mm] 
   + (\alpha_1 |\Psi_\pm|^2+\alpha_2 |\Psi_{\mp}|^2)\Psi_\pm  +\chi\left(\frac{\p\ }{\p X} \mp i\frac{\p\ }{\p Y}\right)^2\Psi_\mp,
    \label{eq:S1}
\end{eqnarray*}
where $X$ and $Y$ are spatial coordinates, $T$ is time, $m^*$ is the effective mass, constants $\alpha_{1,2}$ characterize polariton-polariton interactions, $\chi=\hbar^2(m_\mathrm{TE}^{-1} - m_\mathrm{TM}^{-1})/4$ is a parameter describing TE-TM splitting.  Next, we introduce a normalized spatial coordinates $x=X/\ell$ and $y=Y/\ell$, where $\ell$ is a characteristic scale. Expecting formation of solitons with characteristic widths of order $\sim 10~\mu$m (which is consistent with experiments on bright polariton solitons in GaAs semiconductor microcavities \cite{dark13}), one can reasonably choose  $\ell=1~\mu$m. The unit energy reads ${\cal E} = \hbar^2/(m^*\ell^2)$, and time is normalized as $T= t/\tau$, where   $\tau= \hbar/{\cal E}$. Spin-orbit coupling coefficient $\beta$ in  Eq.~(\ref{eq:full}) of the main text is defined as  $\beta=\chi m^*/{\hbar^2}$, and the Zeeman lattice is connected to the spatially modulated magnetic field as $\Omega(x)=g\mu_B/(2{\cal E})B(X)$. Finally, using normalization of wavefunctions in the form $\Psi_\pm = \sqrt{{\cal E}/\alpha_1}\, \psi_\pm$, one can arrive at Eq.~(\ref{eq:full}) of the main text (in the conservative limit).

\section{(In)stability of dipole solitons}
\label{app:2}
In the conservative limit ($\gamma=0$, $H_\pm=0$), the dimensionless Gross-Pitaevskii equations (\ref{eq:full}) from the main text can be derived as the Hamiltonian equations of motion 
\begin{equation}
    i\partial_t{\psi_+} = \frac{\delta E}{\delta \psi_+^*}, \quad  i\partial_t{\psi_-} = \frac{\delta E}{\delta \psi_-^*}, 
\end{equation}
starting from the Hamiltonian (energy) functional $E = E_{kin}+E_{SOC}+E_{int}+E_Z$, where
\begin{equation}
E_{kin} = \frac{1}{2} \iint dxdy\, (|\nabla\psi_+|^2 + |\nabla\psi_-|^2) 
\end{equation}
is the kinetic term, 
\begin{eqnarray*}
E_{SOC} = -\beta \iint dxdx\,  \left(-\py \psi_+^*\py \psi_- - 2i\px \psi_+^*\py\psi_-  \right.\nonumber\\
\left. + \px\psi_+^*\px\psi_-\right) + \textrm{c.c.}
\end{eqnarray*}
is the term emerging from the spin-orbit interactions, and $E_{int}$ and $E_Z$ are the terms that take into account polariton interactions and inhomogeneous Zeeman splitting:
\begin{eqnarray*}
E_{int} &=& \frac{1}{2}\iint dxdy\, (|\psi_+|^4 + 2\sigma  |\psi_+|^2 |\psi_-|^2 + |\psi_-|^4),\\
E_Z  &=&  \iint dxdy\, \Omega(x) (|\psi_+|^2 - |\psi_-|^2).
\end{eqnarray*}
For $y$-independent soliton solutions $\psi_\pm = e^{-i\mu t}u_\pm(x)$ with real stationary wavefunctions $u_\pm(x)$ the SOC term simplifies to $E_{SOC}=-2\int dx \px u_+ \px u_-$. Assuming that for the dipole solitons, similar to those shown in Fig.~\ref{fig:deep}(b,c) of the main text, the SOC term is mainly determined by the region where wavefunctions $u_+$ and $u_-$ overlap, we readily conclude that for the in-phase dipole shown in Fig.~4(b) one has $\px u_+ \px u_-<0$ in the overlap region, while for an out-of-phase dipole $\px u_+ \px u_->0$ in the overlap region. Thus, for the positive SOC coefficient $\beta>0$ the out-of-phase dipoles are energetically more preferable and have more chances to be dynamically stable than in-phase dipoles. \textit{ Vice versa}, for negative $\beta<0$ the in-phase dipoles are expected to be more stable than the out-of-phase ones.

\end{document}